\documentclass[twocolumn,doublespace,showkeys,showpacs,superscriptaddress]{IEEEtran}

\usepackage{graphicx,epic,eepic,epsfig,amsmath,latexsym,amssymb,verbatim,subfigure,color}
\usepackage{theorem}

\newtheorem{definition}{Definition}
\newtheorem{proposition}[definition]{Proposition}
\newtheorem{lemma}[definition]{Lemma}

\newtheorem{theorem}[definition]{Theorem}

\def\squareforqed{\hbox{\rlap{$\sqcap$}$\sqcup$}}
\def\qed{\ifmmode\squareforqed\else{\unskip\nobreak\hfil
\penalty50\hskip1em\null\nobreak\hfil\squareforqed
\parfillskip=0pt\finalhyphendemerits=0\endgraf}\fi}
\def\endenv{\ifmmode\;\else{\unskip\nobreak\hfil
\penalty50\hskip1em\null\nobreak\hfil\;
\parfillskip=0pt\finalhyphendemerits=0\endgraf}\fi}

\mathchardef\ordinarycolon\mathcode`\:
\mathcode`\:=\string"8000
\def\vcentcolon{\mathrel{\mathop\ordinarycolon}}
\begingroup \catcode`\:=\active
  \lowercase{\endgroup
  \let :\vcentcolon
  }

\newcommand{\nc}{\newcommand}
\nc{\rnc}{\renewcommand}
\nc{\beq}{\begin{equation}}
\nc{\eeq}{{\end{equation}}}
\nc{\beqa}{\begin{eqnarray}}
\nc{\eeqa}{\end{eqnarray}}
\nc{\lbar}[1]{\overline{#1}}
\nc{\bra}[1]{\langle#1|}
\nc{\ket}[1]{|#1\rangle}
\nc{\ketbra}[2]{|#1\rangle\!\langle#2|}
\nc{\braket}[2]{\langle#1|#2\rangle}
\nc{\proj}[1]{| #1\rangle\!\langle #1 |}
\nc{\avg}[1]{\langle#1\rangle}
\nc{\Rank}{\operatorname{Rank}}
\nc{\smfrac}[2]{\mbox{$\frac{#1}{#2}$}}
\nc{\Tr}{\operatorname{Tr}}
\nc{\tr}{\operatorname{Tr}}
\nc{\id}{\operatorname{id}}
\nc{\1}{\openone}
\nc{\ox}{\otimes}
\nc{\dg}{\dagger}
\nc{\dn}{\downarrow}
\nc{\cA}{{\cal A}}
\nc{\cB}{{\cal B}}
\nc{\cC}{{\cal C}}
\nc{\cD}{{\cal D}}
\nc{\cE}{{\cal E}}
\nc{\cF}{{\cal F}}
\nc{\cG}{{\cal G}}
\nc{\cH}{{\cal H}}
\nc{\cI}{{\cal I}}
\nc{\cJ}{{\cal J}}
\nc{\cK}{{\cal K}}
\nc{\cL}{{\cal L}}
\nc{\cM}{{\cal M}}
\nc{\cN}{{\cal N}}
\nc{\cO}{{\cal O}}
\nc{\cP}{{\cal P}}
\nc{\cR}{{\cal R}}
\nc{\cS}{{\cal S}}
\nc{\cT}{{\cal T}}
\nc{\cX}{{\cal X}}
\nc{\cY}{{\cal Y}}
\nc{\cZ}{{\cal Z}}
\nc{\supp}{{\operatorname{supp}}}
\nc{\var}{\operatorname{var}}
\nc{\rar}{\rightarrow}
\nc{\lrar}{\longrightarrow}
\nc{\polylog}{\operatorname{polylog}}

\def\e{\epsilon}

\nc{\RR}{{{\mathbb R}}}
\nc{\CC}{{{\mathbb C}}}
\nc{\FF}{{{\mathbb F}}}
\nc{\NN}{{{\mathbb N}}}
\nc{\ZZ}{{{\mathbb Z}}}
\nc{\PP}{{{\mathbb P}}}
\nc{\QQ}{{{\mathbb Q}}}
\nc{\UU}{{{\mathbb U}}}
\nc{\EE}{{{\mathbb E}}}
\nc{\Icoh}{{I^{\rm coh}}}
\nc{\Qca}{{Q_{\rm ss}}}
\nc{\Qcaa}{{Q^{(1)}_{\rm ss}}}
\nc{\Dcaa}{{D^{(1)}_{{\rm ss}\rightarrow}}}
\nc{\Dca}{{D_{{\rm ss}\rightarrow}}}

\rnc{\1}{{{\mathbb I}}}

\nc{\be}{\begin{equation}}
\nc{\ee}{{\end{equation}}}
\nc{\bea}{\begin{eqnarray}}
\nc{\eea}{\end{eqnarray}}
\nc{\<}{\langle}
\rnc{\>}{\rangle}
\nc{\Hom}[2]{\mbox{Hom}(\CC^{#1},\CC^{#2})}
\nc{\rU}{\mbox{U}}

\begin{document}
\title{The quantum capacity with symmetric side channels}

\author{Graeme Smith, John A. Smolin and Andreas Winter
\thanks{Graeme Smith was at the Institute for Quantum Information, Caltech 107--81,
    Pasadena, CA 91125, USA, and is currently at the IBM T.J. Watson Research Center, Yorktown Heights, NY 10598, USA.}
\thanks{John A. Smolin is at the IBM T.J. Watson Research Center, Yorktown Heights, NY 10598, USA}
\thanks{Andreas Winter is at the Department of Mathematics, University of Bristol,
Bristol BS8 1TW, United Kingdom }
\thanks{Graeme Smith received financial support from the US NSF
(project PHY-0456720), and NSERC of Canada.  John Smolin acknowledges the support of ARO contract DAAD19-01-C-0056.
Andreas Winter received support from the U.K. EPSRC via ``QIP IRC'' and the European Commission under
project ``QAP'' (contract IST-2005-15848), as well as a University of Bristol Research Fellowship.}
}

\maketitle
\date{\today}

\begin{abstract}
  We present an upper bound for the quantum channel capacity that is
  both additive and convex.
  Our bound can be interpreted as the capacity of a channel for
  high-fidelity quantum communication  when assisted by a family of channels
  that have no capacity on their own.  This family of assistance channels, which we
  call symmetric side channels, consists 
  of all channels mapping symmetrically to their output and environment.
  The bound seems to be quite tight, and for degradable quantum channels 
  it coincides with the unassisted channel capacity.
  Using this symmetric side channel capacity,
  we find new upper bounds on the capacity of the depolarizing channel.
  We also briefly indicate an analogous notion for distilling entanglement
  using the same class of (one-way) channels, yielding one of the
  few entanglement measures that is monotonic under local operations with one-way classical communication (1-LOCC), but
  not under the more general class of local operations with classical communication (LOCC).

\end{abstract}

\section{Introduction}
The archetypical problem in information theory is finding the capacity of a
noisy channel to transmit messages with high fidelity.
Already in \cite{Shannon48}, Shannon provided a simple formula
for the capacity of a discrete memoryless channel,
with single-letter capacity formulas of more general channels
to follow later (see e.g.~\cite{CoverThomas}).

The status of the quantum channel capacity question is not
nearly as nice. While there has recently been 
significant progress towards finding the quantum
capacity of a quantum channel \cite{Lloyd97,Shor02,D03},
the resulting expressions cannot be evaluated in any
tractable way, with the exception of some very special 
channels (e.g., the capacity of the amplitude-damping \cite{GF04}, 
dephasing \cite{Rains99} and erasure \cite{BDS97} channels are known, most others are not).
In fact, there are several capacities that can be defined for a quantum channel, depending on 
what type of information is to be sent (e.g., quantum or classical) and what sort of resources are allowed 
to accomplish transmission (e.g., free entanglement, two-way classical communication, etc.).
So far only two of these capacities seem to admit single-letter formulas:
the entanglement-assisted capacity \cite{BSST02,AC97} and the environment-assisted quantum capacity \cite{SVW05,W05}.  
The multi-letter formulas available for the other capacities, including the quantum capacity,
provide, at best, partial characterizations.

For instance, it was shown in \cite{SN96,Lloyd97,Shor02,D03} that
the capacity for noiseless quantum communication of a quantum channel $\cN$ is given by 
\begin{equation}
\label{Eq:QuantumCapacity}
  Q(\cN)=\lim_{n\rightarrow\infty} \frac{1}{n}
                                      \max_{\ket{\phi}_{A(A^\prime)^{\ox n}}} I(A\rangle B^{\ox n})_{\omega_{AB^{\ox n}}}. 
\end{equation}
In this expression, $\cN$ is a quantum channel mapping quantum states on the vector space $A'$ to states on the space $B$, 
and $\ket{\phi}_{A(A')^{\ox n}}$ is a pure quantum state on $n$ copies of $A'$ together with a reference system $A$.
The state $\omega_{AB^{\ox n}} = \id\ox \cN^{\ox n}(\proj{\phi}_{A(A^\prime)^{\ox n}})$, is the state that results when the $n$
copies of $A'$ are acted on by $n$ copies of the channel $\cN$.  Finally,   
$I(A\rangle B^{\ox n})_{\omega_{AB^{\ox n}}} = S(\omega_{B^{\ox n}}) - S(\omega_{A B^{\ox n}})$
is known as the {\em coherent information} \cite{SN96}, which is defined in 
terms of the von Neumann entropy $S(\rho) = -\Tr \rho\log \rho$.  In order to evaluate this regularized 
formula one would have to perform an
optimization over an infinite number of variables, making a numerical approach essentially impossible.
Furthermore, it is known that the limit on
the right is in general strictly larger than the corresponding single-letter 
expression \cite{DSS98,SS96,SmithSmo0506}: there are channels, $\cN$, for which
\begin{equation}
  Q^{(1)}(\cN) := \max_{\ket{\phi}_{AA^\prime}} I(A\rangle B)_{\omega_{AB}} < Q(\cN).
\end{equation}

In the absence of an explicit formula for the quantum capacity,
it is desirable to find upper and lower bounds for Eq.~(\ref{Eq:QuantumCapacity}).
Unfortunately, most known bounds are as difficult to evaluate
in general as Eq.~(\ref{Eq:QuantumCapacity}).  Examples of upper bounds 
that {\em can} be easily evaluated, at least in some special cases,
are given by the no-cloning based arguments of \cite{BDEFMS98,Cerf00},
the semi-definite programming bounds of Rains \cite{Rains99,Rains-PPT} and the
closely related relative entropy of entanglement \cite{VP97}.
None of these is expected to  be particularly tight---the last two are
also upper bounds for the capacity assisted by two-way classical 
communication (which can be much larger than one-way),
whereas the first is based solely on reasoning about where the
channel's capacity must be zero.  As such, it would be useful to
find new upper bounds for the quantum capacity that are both free
of regularization and fundamentally one-way.
In the following we present just such a bound.

Inspired by the fact that allowing free forward classical communication
does not increase the quantum channel capacity \cite{BarnumKN00}, we will consider the
capacity of a quantum channel assisted by the use of a quantum channel
that maps symmetrically to the receiver (Bob) and the environment (Eve).
Such assistance channels, which we call {\em symmetric side channels}, can be
used for forward classical communication but are apparently somewhat stronger.
They can, however,
immediately be seen to have zero quantum capacity, so that while the assisted
capacity we find may in general be larger than the usual quantum capacity,
one expects that it will provide a fairly tight upper bound. In particular, 
the {\em symmetric side channel capacity} (ss-capacity) we find will not be an upper
bound for the capacity assisted by two-way classical communication.

The expression we find for the assisted capacity, which we'll 
call $\Qca$, turns out to be
much easier to deal with than Eq.~(\ref{Eq:QuantumCapacity}) and has several nice properties.  
Most importantly, our expression is free of the regularization present
in so many quantum capacity formulas.  We will also see that $\Qca$ is
convex, additive, and that it is equal to $Q$ for the family 
of degradable channels \cite{DS03}.
We will use these properties to find upper bounds on
$\Qca$ of the depolarizing channel which, in turn, 
will give a significant improvement over known bounds for its
unassisted capacity.

It should be emphasized that we have not found an upper bound on the dimension 
of the side channel needed to attain the assisted capacity, which 
in general prevents us from evaluating  $\Qca$ explicitly or even
numerically. While we 
cannot rule out such a bound, the arguments we use to establish 
several of $\Qca$'s nice properties rely explicitly on the availability 
of an unbounded dimension. This suggests that dealing
with an assistance channel of unbounded dimension may be the 
price we pay for such desirable properties as additivity and convexity,
which is reminiscent of the findings of \cite{BHLS03,CW04}.

\section{Preliminaries}
In this section, we collect the definitions of important concepts and quantities, as well as describing some of their properties.

We will mainly be concerned with finite-dimensional quantum systems.  The state of a $d$ dimensional system is described
by a {\em density operator} (or {\em density matrix}), which is a trace one linear operator on the complex vector space $\CC^d$, 
typically denoted $\rho\in \cB(\CC^d)$, where we have used the notation $\cB(\cH)$ to denote the set of 
bounded linear operators on a space $\cH$. Such a $\rho$ is required to be hermitian, meaning that $\rho = \rho^\dg$ where the hermitian 
conjugate $\dg$ consists of transposition followed by complex conjugation, and positive semidefinite, meaning $\rho \geq 0$.  
Any such $\rho$ has  a spectral decomposition, $\rho = \sum_{i=1}^{d}\lambda_i \proj{\phi_i}$, where $\proj{\phi_i}$ denotes
the projector onto an element $\ket{\phi_i} \in \CC^d$, the $\ket{\phi_i}$ satisfy $\braket{\phi_j}{\phi_i} = \delta_{ij}$, 
and the $\lambda_i$s are nonnegative and sum to one.  A rank one density operator, $\rho = \proj{\phi}$, is called a pure state.
We will often include the pure state and density operator's spaces as subscripts, for example $\rho_A$ denotes a density operator on $A$
and $\ket{\phi}_A \in A$.  

A useful operation on the set of quantum states is the partial trace.  We first define the usual trace of 
a density operator $\rho = \sum_i \lambda_i \proj{\phi_i}$ to be $\Tr\rho = \sum_i \lambda_i$.  
If $\rho_{AB}$ is a density operator on the tensor product 
of $A$ and $B$, $A\otimes B$, we define the partial trace over $B$, denoted $\Tr_B$ as the unique linear operation satisfying
\begin{equation}
\Tr(\proj{\psi} (\Tr_B \rho_{AB}) ) = \Tr((\proj{\psi}\otimes \1_B)\rho_{AB})
\end{equation}
for all $\ket{\psi}\in A$, and where we have let $\1_B$ be the identity on $B$.  Physically, the partial trace over $B$ may be thought of
as discarding the $B$ system.  The resulting state on $A$ is referred to as the reduced state on $A$.  
Given an state $\rho_{AB}$, we will often use subscripts to denote a reduced state, for
example $\rho_A = \Tr_B \rho_{AB}$.  We will often be concerned with quantum states on the 
tensor product of many copies of the same space, where we will use the notation 
$A^{\otimes n} = \stackrel{n \  {\rm times}}{\overbrace{A \otimes {\dots} \otimes A}} $, 
and occasionally $A^n = A^{\otimes n}$.

Given two states, $\rho$ and $\sigma$, a natural measure of their similarity is the fidelity, 
\begin{equation}
F(\rho,\sigma) = \Tr\sqrt{\sqrt{\sigma} \rho\sqrt{\sigma}},
\end{equation}
which is equal to $1$ if the states are identical and $0$ if they are orthogonal.  Another useful 
measure of their similarity is the trace distance, defined as
\begin{equation}
D(\rho,\sigma) = \frac{1}{2}\Tr |\rho-\sigma|,
\end{equation}
where $|A| = \sqrt{A^\dg A}$.  These two measures are related \cite{FvG99} according to
\begin{equation}\label{Eq:Fuchs}
1-F(\rho,\sigma) \leq D(\rho,\sigma) \leq \sqrt{1 - F(\rho,\sigma)^2}.
\end{equation}

The physical operations that can be applied a quantum state are {\em completely positive trace preserving} (CPTP) linear maps 
from $\cB(\cH_1)$ to $\cB(\cH_2)$, where $\cH_1$ and $\cH_2$ are the input and output spaces, respectively.  A positive 
linear map, $\cN$, satisfies the requirement $\cN(\rho) \geq 0$ for every $\rho \geq 0$.  In addition, a linear map with input 
space $\cH_1$ and output space $\cH_2$ can be extended to a map from $\cH_3\otimes \cH_1$ to $\cH_3 \otimes \cH_2$, where $\otimes$
denotes a tensor product of the spaces, by choosing the extended map to act as the identity on $\cH_3$.  If the extended map, which 
we will denote $\id_{\cH_3} \ox \cN$, is positive for any choice of $\cH_3$, the map $\cN$ is  called {\em completely positive}.  Together
with the trace-preserving requirement, demanding complete positivity ensures that CPTP maps are the most general class of linear operations  
mapping density operators to density operators.  Due to the Stinespring dilation theorem \cite{Stine}, 
a CPTP map (or {\em quantum channel}) $\cN$,
with input space $A$ and output space $B$ can always be represented as an isometric embedding of $A$ into $B\otimes E$ for some 
environment space $E$, followed by a partial trace over $E$.  In other words, there will be an isometry $U: A \rightarrow B\otimes E$, 
satisfying $U^\dg U = \id_A$, such that $\cN(\rho) = \Tr_E U\rho U^\dg$. Sometimes the isometry
corresponding to a channel $\cN$ will be called $U_{\cN}$.  This dilation, of which
we shall make free use, is unique up to unitary equivalences of $E$.

There is another representation of a quantum channel is in terms of its Kraus decomposition.  Any quantum channel with input space $A$
and output space $B$ can be expressed as
\begin{equation}
\cN(\rho) = \sum_k A_k \rho A_k^\dg,
\end{equation}
where $A_k$ are linear maps from $A$ to $B$ with $\sum_k A_k^\dg A_k = {\1}_B$, and ${\1}_B$ is the identity on $B$.  In contrast to
$\1_B$, which is an operator on the vector space $B$, we denote the identity channel on $\cB(B)$ as $\id_B$, which acts according
to $\id_B(\rho) = \rho$ for all $\rho\in \cB(B)$.

A channel of particular interest is the depolarizing channel, which maps a two-dimensional space (or, {\em qubit}) to a two-dimensional
space.  This channel is the quantum analogue of the binary symmetric channel.  For any qubit density operator, 
$\rho \in \cB(\CC^2)$, the depolarizing channel with error probability $p$ acts as
\begin{equation}
\cN_p(\rho) = (1-p)\rho + \frac{p}{3}X\rho X + \frac{p}{3}Y\rho Y + \frac{p}{3}Z\rho Z,
\end{equation}
where $X$, $Y$, and $Z$ are the Pauli matrices, 

\begin{eqnarray}
X &=& \left(\begin{matrix}0 & 1\\ 1 & 0\end{matrix}\right) \\
Y &=& \left(\begin{matrix}0 & -i\\ i & 0 \end{matrix}\right)\\
Z &=& \left(\begin{matrix}1 & 0\\ 0 & -1 \end{matrix}\right). 
\end{eqnarray}
Even the capacity of this relatively simple quantum channel is unknown.  In Section \ref{sec:application} we will find upper bounds
on this capacity.

The von Neumann entropy of a density operator $\rho$ on a space $A$ is given by $S(\rho) = -\Tr\rho \log \rho$.  We will 
often use the notation $S(A)_{\rho}$ to denote the entropy of a state $\rho$ on a space $A$ and, 
when it is clear to which state we refer, we will also simply write $S(A)$.  The {\em coherent information of A given B}
of a bipartite state $\rho_{AB} \in \cB(A\ox B)$ is defined as
\begin{equation}
I(A\rangle B)_{\rho_{AB}} = S(\rho_B) - S(\rho_{AB}),
\end{equation}
or equivalently, $I(A\rangle B)_{\rho_{AB}} = S(B)_{\rho_B} - S(AB)_{\rho_{AB}}$.  As with the entropy, when there is 
no ambiguity as to which state is being discussed, we will simply write $I(A\rangle B) = S(A) -S(AB)$.
The coherent information satisfies a quantum data-processing inequality
with respect to processing on the $B$ system, meaning that for any state $\rho_{AB}$ and channel, $\cN$, mapping $B$ to $C$,
\begin{equation}\label{Eq:DataProcessing}
I(A\rangle B)_{\rho_{AB}} \geq I(A\rangle C)_{\id_A \ox \cN (\rho_{AB})}.
\end{equation}
This data processing inequality is a simple consequence of the strong subadditivity of von Neumann entropy \cite{LiebRuskai},
and was first pointed out by \cite{SN96}.  
The failure of the analogous data processing inequality 
on the $A$ system \cite{BNS98,SS96} is closely related to the need for a regularization in the formula for the quantum 
channel capacity in Eq.~(\ref{Eq:QuantumCapacity}).

A useful property of the von Neumann entropy is that is continuous---two states which 
are close in terms of trace distance have entropies which are correspondingly close.  More
specifically, Fannes has shown \cite{Fannes73} that if $\rho$ and $\sigma$ are states on a 
$d$-dimensional space with trace distance $D(\rho,\sigma) \leq 1/e$, then
\begin{equation}
|S(\rho)-S(\sigma)| \leq D(\rho,\sigma)\log d - D(\rho,\sigma)\log\left(D(\rho,\sigma)\right).
\end{equation}
If we do not require $D(\rho,\sigma) \leq 1/e$, we have a slightly looser bound of
\begin{equation}
|S(\rho)-S(\sigma)| \leq D(\rho,\sigma)\log d + \frac{\log e}{e}.
\end{equation}
In light of the relationship between fidelity and trace distance expressed in 
Eq.~(\ref{Eq:Fuchs}), we also have the relation
\begin{equation}\label{Eq:FannesFidelity}
|S(\rho)-S(\sigma)| \leq \sqrt{1-F(\rho,\sigma)}\log d + \frac{\log e}{e},
\end{equation}
which we will find useful in proving the converse of our coding theorem below.

Finally, we will occasionally use the {\em quantum mutual information}, 
\begin{equation}
I(A;B)_{\rho_{AB}} = S(A)_{\rho_A} + S(B)_{\rho_B} - S(AB)_{\rho_{AB}},
\end{equation}
which derives an operational meaning from its role in the single-letter formula for the
entanglement assisted capacity \cite{BSST02}.

\section{Definitions and Properties of Capacities}
\label{sec:defi}

\subsection{Unassisted Quantum Capacity}

Before studying the symmetric side channel assisted capacity, we first review the usual, unassisted, quantum
capacity problem.  In this scenario, illustrated in Fig. 1, our sender and receiver are 
given access to asymptotically many uses of a quantum channel: $\cN^{\ox n}$.  If the input space of
$\cN$ is $A$ and the output space $B$, our goal is to find a subspace $C \subset A^{\otimes n}$ and a 
decoding operation $\cD_n : \cB(B^{\otimes n}) \rightarrow \cB(C)$ such that every state $\ket{\psi}\in C$ can be decoded with high
fidelity after it is sent through the channel:
\begin{equation}
\cD_n \circ \cN^{\ox n}(\proj{\psi}) \approx \proj{\psi}.
\end{equation}
Of course, our goal is to find the largest possible code $C$.

More formally, we say a rate $R$ is achievable if for every $\e >0$ and sufficiently large $n$, there is a 
code $C_n \subset A^{\ox n}$ with $\log \dim C_n \geq Rn$ and a decoding operation $\cD_n:\cB(B^{\ox n}) \rightarrow \cB(C_n)$ 
such that for all $\ket{\psi} \in C_n$, the fidelity 
\begin{equation}
F\left(\proj{\psi}, \cD_n \circ \cN^{\ox n}(\proj{\psi})\right) \geq 1 - \e.
\end{equation}
The {\em capacity} of $\cN$ is defined to be the largest such achievable rate.

The best known strategy for generating good quantum codes is based on a random coding argument \cite{Shor02,D03}.
Given a channel $\cN$ mapping $A^\prime$ to $B$ and a state $\ket{\phi}_{AA^\prime}$,  the reduced 
state $\phi_{A^\prime} = \Tr_A \proj{\phi}_{AA^\prime}$ provides a prescription for generating good codes with rates 
up to the coherent information,
\begin{equation}
R = I(A\rangle B)_{(\1_A \ox \cN)(\proj{\phi}_{AA^\prime})}.
\end{equation}
If one chooses the basis of a blocklength $n$ code by selecting random vectors that are, roughly speaking, 
distributed like $\phi_{A^\prime}^{\ox n}$, as long as the rate of the code is no more than this coherent information, 
it will with high probability allow high fidelity transmission.  

As it turns out, when one evaluates the coherent information that can be generated with $m$ uses of a channel, it will 
in some cases exceed $m$ times the maximum coherent information that can be generated with one copy.  This means 
that by using codes that are not chosen to resemble some i.i.d. input state, but rather whose distribution is correlated 
across several copies of the channel, it is possible to find better codes. 
All known examples of
this effect occur in channels for which the single-letter coherent information is either zero or very small, where it seems to be
rather generic.  While some progress was made in \cite{SmithSmo0506}, there is still no systematic understanding of how to generate
non-i.i.d. high performance codes.

\begin{figure}[htbp]
\includegraphics[width=8cm]{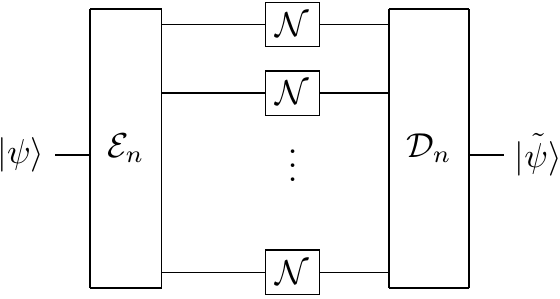}
\caption{The unassisted quantum capacity problem.  Given $n$ uses of a quantum channel, $\cN:\cB(A')\rightarrow \cB(B)$
 we would like to find a quantum code $C_n \subset (A')^{\ox n} $ such that every $\ket{\psi}\in C_n$ can be decoded with high fidelity
after being sent through $\cN^{\ox n}$.  The rate of $C_n$ is defined as $R = \frac{1}{n}\log \dim C_n$, and the optimal such rate is
called the quantum capacity.  The best known expression for the quantum capacity is the multi-letter formula 
in Eq.~(\ref{Eq:QuantumCapacity}). }
\end{figure}

\subsection{Symmetric Side Channel Assisted Capacity}

We now turn to our assisted quantum capacity problem.  First let $W_d \subset \top \ox \perp$ be the
$d(d+1)/2$-dimensional symmetric subspace between 
$d$-dimensional spaces $\top$ and $\perp$.  $W_d$ is spanned by the following basis labeled by $i,j \in \{1,\dots,d\}$ with $i\leq j$:
\begin{eqnarray}
\ket{(i,j)} & = & \frac{1}{\sqrt{2}}\left( \ket{i}\ket{j}+\ket{j}\ket{i}\right) {\ \ {\rm for}\  i \neq j }\\
& = & \ket{i}\ket{i} \ \ {\rm for} \ i = j.
\end{eqnarray}
Now, we let $V_d: \CC^{d(d+1)/2} \rightarrow W_d$ be an isometry which maps a basis of $\CC^{d(d+1)/2}$ to these $\ket{(i,j)}$ in
some order.  The \emph{$d$-dimensional symmetric side channel} is defined to be the channel mapping $\cB(\CC^{d(d+1)/2})$ to $\cB(\top)$
that is obtained by applying $V_d$ followed by the partial trace over $\perp$:
\begin{equation}\label{eq:clone}
\cA_d(\rho) = \Tr_{\perp}V_d\rho V_d^\dg.
\end{equation}

Because $\cA_d$ maps symmetrically between its output ($\top$) and environment ($\perp$), its quantum capacity will turn out to be zero.
As a result, one would expect that allowing $\cA_d$ as a free resource to be used along with some channel $\cN$, 
the resulting assisted capacity would provide a reasonably tight upper bound for the unassisted capacity of $\cN$.  Furthermore, when we
define such an assisted capacity, we will find that it is much better behaved than the unassisted capacity seems to be.

\begin{figure}[htbp]
\label{fig:assisted}
\includegraphics[width=8cm]{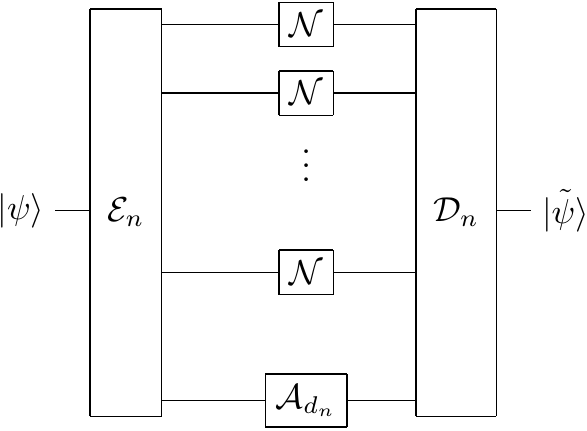}
\caption{The quantum capacity with symmetric assistance.  Given $n$ uses of  a quantum channel, $\cN:\cB(A')\rightarrow \cB(B)$, 
we also now have free access to a zero-capacity symmetric side channel with arbitrary output dimension, 
$\cA_{d_n}:\cB(\CC^{d_n(d_n+1)/2})\rightarrow \cB(\CC^{d_n})$.  Our goal is to find the highest rate subspace of the input spaces 
$(A')^{\ox n} \ox \CC^{d_n(d_n+1)/2}$ that still allows high-fidelity reconstruction of every state in the space
 after the channels have been applied.  The best known expression for the capacity in this setting is the single-letter formula
 of Eq.~(\ref{Eq:SSCapacity}).}
\end{figure}

Formally, for a channel $\cN:\cB(\tilde{A})\rightarrow \cB(B)$,  we say that a rate $R$ is \emph{ss-achievable} if for all $\e>0$ and
sufficiently large $n$, there is a dimension $d_n$, a
code $C_n \subset \tilde{A}^{\ox n}\ox W_{d_n}$ with $\log\dim C_n \geq Rn$, 
and a decoding operation $\cD_n: \cB(B^{\otimes n}\ox \CC^{d_n})$ such that
for all states $\ket{\psi}\in C_n$,  the reconstructed state 
$\cD_{n}\bigl[ (\cN^{\ox n }\ox \cA_{d_n})\proj{\psi} \bigr]$ has a fidelity of at least $1-\e$ with 
the original state $\proj{\psi}$.  The {\em ss-capacity}, which we will denote
by $\Qca(\cN)$, is defined as the supremum of all ss-achievable rates.

Note that assistance by the symmetric channels includes free use of
classical communication, as the dephasing operation
$\ket{x}\longrightarrow\ket{x}\ket{x}$ is obtained by restricting $\cA_d$
to a subspace.

We are now in a position to introduce a quantity that will 
play a central role in our study of the ss-capacity.  
Letting $\cN: \cB(\tilde{A})\rightarrow \cB(B)$ 
be a channel, we define $\Qcaa(\cN)$ to be
the supremum over all states $\proj{\phi}_{A\tilde{A}\top\perp}$ 
that are invariant under the permutation of $\top$ and $\perp$,
of the coherent information of $A$
given $B\top$, evaluated after the $\tilde{A}$ register of 
$\phi$ is acted on by $\cN$.  That is, we let 
\begin{align}
  \label{Eq:omega}
  \omega_{AB\top\perp} &= (\id_{A\top\perp}\ox\cN)\phi_{A\tilde{A}\top\perp}, \\
  \label{Eq:Q1}
  \Qcaa(\cN)           &= \sup_{\phi_{A\tilde{A}\top\perp}} I(A\rangle B\top)_\omega
                        = \sup_d\,                  Q^{(1)}\bigl( \cN\ox\cA_d \bigr),
\end{align}
where the supremum is over all pure states $\phi_{A\tilde{A}\top\perp}$ invariant
under the swap $\top\leftrightarrow\perp$ of $\top$ and $\perp$. The rightmost, alternative,
expression for $\Qcaa(\cN)$ is seen as follows.  On the one hand, 
for every state $\ket{\phi} \in A\tilde{A}W_d$, 
$(\1_{A\tilde{A}}\ox V_d)\ket{\phi}$ is a state on $A\tilde{A}\top\perp$
that is symmetric in $\top\bot$, so that the coherent information of $(\id\ox\cN\ox\cA_d)\phi_{A\tilde{A}W_d}$ is
exactly $I(A\rangle B\top)$. On the other hand, if we have a pure 
state $\phi_{A\tilde{A}\top\perp}$ that is invariant under the exchange of 
$\top$ and $\perp$, it must be an eigenvector
of the swap operator with eigenvalue $1$ or $-1$. In the latter case
we can extend $\top$ and $\perp$ with a qubit and tensor a singlet
onto $\ket{\phi}$---this doesn't change the coherent information
but results in a vector $\ket{\phi}$ which is invariant under swapping
$\top$ and $\perp$. As a result, $\tr_{A\tilde{A}}\phi$ is supported on the
symmetric subspace of $\top\perp$ and we can present $\ket{\phi}$
as the image of a pure state under some $\1_{A\tilde{A}}\ox V_d$.

\medskip
For later use, we start by deriving a different formula for $\Qcaa$.
\begin{lemma}
  \label{lemma:asymmetric}
  For any channel $\cN$ with Stinespring dilation $U_{\cN}: A \rightarrow BE$,
  \begin{equation}
    \label{eq:asym}
    \Qcaa(\cN) = \sup_{\rho_{A\tilde{A}F}}
                    \frac{1}{2}\bigl[ I(A\rangle BF)_{\omega} - I(A\rangle EF)_{\omega} \bigr],
  \end{equation}
  with respect to the state
  $\omega_{ABEF} = (\1_{AF}\ox U_{\cN})\rho(\1_{AF}\ox U_{\cN})^\dagger$.
\end{lemma}
\begin{proof}
  We may think of $\rho_{A\tilde{A}F}$ as the reduced state
  $\tr_{F'}\phi_{A\tilde{A}FF'}$ of a pure state $\ket{\phi}$,
  and look at the information quantities in the lemma w.r.t.~the
  state $(\1_{AFF'}\ox U_{\cN})\ket{\phi}$.
  Then, it is an elementary identity that $I(A\rangle EF) = -I(A\rangle BF')$,
  and in the r.h.s.~of Eq.~(\ref{eq:asym}) the expression becomes
  \[
    \frac{1}{2}\bigl[ I(A\rangle BF) + I(A\rangle BF') \bigr].
  \]
  Notice that if $\phi$ is symmetric under swapping $F$ and $F'$, this is
  equal to $I(A\rangle BF)$.

  In general, we can, with $\top = FG$ and $\perp = F'G'$ (where $G$ and $G'$
  label qubit registers), define
\begin{eqnarray}\nonumber
\ket{\varphi}_{A\tilde{A}\top\perp} &=&  \frac{1}{\sqrt{2}} \ket{\phi}_{A\tilde{A}FF'}\ket{01}_{GG'}\\
 & & + \frac{1}{\sqrt{2}}(\1_{A\tilde{A}}\ox{\rm SWAP}_{FF'})\ket{\phi}_{A\tilde{A}FF'}\ket{10}_{GG'},\nonumber
\end{eqnarray}
where ${\rm SWAP}_{FF'}\ket{i}_F\ket{j}_{F'} = \ket{j}_F\ket{i}_{F'}$ is a unitary that permutes $F$ and $F'$.  
Then, with respect to the state $\Omega_{AB\top\perp}=(\id_{A\top\perp}\ox\cN)\varphi$,
  \[
    \frac{1}{2}\bigl[ I(A\rangle BF) + I(A\rangle BF') \bigr]_\omega = I(A\rangle B\top)_\Omega,
  \]
  and we are done.
\end{proof}

\medskip
It will turn out that $\Qcaa(\cN)$ is exactly the 
ss-capacity of $\cN$, as the following theorem shows. 
\begin{theorem}
  \label{thm:main}
  For all channels $\cN$,
\begin{equation}\label{Eq:SSCapacity}
\Qca(\cN) = \Qcaa (\cN) = \sup_{\phi_{A\tilde{A}\top\perp}} I(A\rangle B\top)_\omega 
\end{equation}
with $\omega_{AB\top\perp} = (\id_{A\top\perp}\ox\cN)\phi_{A\tilde{A}\top\perp}$ and where 
the optimization is over all $\phi_{A\tilde{A}\top\perp}$
invariant under permuting $\top$ and $\bot$.
\end{theorem} 
We will prove this with the following two lemmas.

\begin{lemma}
  \label{Lemma:SubAdd}
  $\Qcaa$ is additive; that is,
  $\Qcaa(\cN_1\ox \cN_2) = \Qcaa(\cN_1) + \Qcaa(\cN_2)$
  for arbitrary channels $\cN_1$ and $\cN_2$.
\end{lemma}
\begin{proof}
  We use the previous lemma, and observe, for a state
  $\rho_{A\tilde{A}_1\tilde{A}_2F}$, and
  \[
    \omega_{AB_1E_1B_2E_2F} = (\1_{AF}\ox U_{\cN_1} \ox U_{\cN_2})
                                \rho(\1_{AF}\ox U_{\cN_1} \ox U_{\cN_2})^\dagger,
  \]
  the identity (with respect to $\omega$)
  \begin{eqnarray}\nonumber    
 I(A\rangle B_1B_2F) - I(A\rangle E_1E_2F) = \ \ \ \ \ \ \  \ \ \ \ \ \ \ \ \ \ \ \ \ \ \ \ \ \ \\
 \nonumber \bigl( I(A\rangle B_1B_2F) - I(A\rangle E_1B_2F) \bigr)\\
          + \bigl( I(A\rangle E_1B_2F)- I(A\rangle E_1E_2F) \bigr).\label{eq:split}
  \end{eqnarray}
  If we introduce new auxiliary systems $F_1 := B_2F$ and $F_2 := E_1F$,
  the above right hand side becomes
  \[
    \bigl( I(A\rangle B_1F_1) - I(A\rangle E_1F_1) \bigr)
            + \bigl( I(A\rangle B_2F_2)- I(A\rangle E_2F_2) \bigr),
  \]
  which is evidently upper bounded by $\Qcaa(\cN_1) + \Qcaa(\cN_2)$,
  while the supremum of the left hand side in Eq.~(\ref{eq:split})
  is $\Qcaa(\cN_1\ox \cN_2)$.
  This shows $\Qcaa(\cN_1\ox \cN_2) \leq \Qcaa(\cN_1) + \Qcaa(\cN_2)$.

  Furthermore, by restricting the optimization in Eq.~(\ref{Eq:Q1}) 
  to states of the form $\phi_{A_1\tilde{A}_1U_1V_1} \ox \phi_{A_2\tilde{A}_2U_2V_2}$
  we see that $\Qcaa(\cN_1\ox \cN_2) \geq \Qcaa(\cN_1) + \Qcaa(\cN_2)$.
\end{proof}

\medskip

Lemma \ref{Lemma:SubAdd} is the key to showing that the ss-capacity has a single-letter formula.  Because this result is 
central to our study, we comment briefly on why it works.  This lemma says that by using $\cN_1$ and $\cN_2$ together with a symmetric 
side channel to generate coherent information, one does no better than if one uses each $\cN_i$ individually to generate ss-assisted coherent
information.  Given a joint input state to $\cN_1\ox \cN_2 \ox \cA_d$, Lemmas \ref{lemma:asymmetric} and 
\ref{Lemma:SubAdd}  give a prescription for generating
an input state for $\cN_1\ox \cA_{d_1}$ by symmetrizing the output and environment of $\cN_2$, and similarly for $\cN_2\ox \cA_{d_2}$.
In fact, the
sum of the coherent informations obtained in this way is at least as much as 
the total coherent information generated with the joint state.  From 
this explanation, we see also see that it is important to allow a large output dimension for our symmetric side channel.

The other ingredient we need is the following multi-letter expression 
for the ss-capacity, which follows by standard arguments
(see, e.g., \cite{D03}).
\begin{lemma}
  \label{Lemma:Regularized}
  The ss-capacity $\Qca$ is given by the regularization of $\Qcaa$: for
  any channel $\cN$,
  \begin{equation}
    \Qca(\cN) = \lim_{n\rightarrow \infty} \frac{1}{n}\Qcaa(\cN^{\ox n}).
  \end{equation}
\end{lemma}
\begin{proof}
  To see that the ss-capacity is no less than the right hand
  side, note that for any $\phi_{A^nB^n\top\perp}$ symmetric under the interchange of $\top$ and $\perp$, the rate 
  $\frac{1}{n}I(A^n\rangle B^n\top)$
  is achievable by the quantum noisy channel coding theorem 
  applied to the channel $\cN^{\ox n}\ox \cA_{d_\top}$
  \cite{Lloyd97,Shor02,D03}.

  To prove the converse, fix $\e$, let $C\subset \tilde{A}^{\otimes n} W_{d_{\top}}$ 
  be an $(n,\e)$-code of rate $R$ making use of a
  symmetric side channel with output dimension $d_\top$ and let 
  $\ket{\phi}_{CD}$ be a state that is maximally entangled between
  the subspace $C$ and a reference system $D$.  Then, with the
  state $\omega = (\id \ox \cN^{\ox n}\otimes \cA_{d_\top})\phi$,
  \begin{align}
    I(D\rangle B^n\top)_\omega 
       &\geq I(D\rangle C)_{(\id \ox \cD_{B^n\top})\omega} \nonumber \\
       &\geq Rn - \frac{2\log e}{e} - 3\log(d_C)\sqrt{\e}\\
        &=     Rn - \frac{2\log e}{e} - 3Rn\sqrt{\e},
  \end{align}
where we have made use of Eq.~(\ref{Eq:FannesFidelity}) twice.
  As a result, we find
  $R \leq (1-3\sqrt{\e})^{-1}\left( \frac{1}{n}\Qcaa(\cN^{\ox n}) + \frac{2\log e}{ne} \right)$, which completes the proof.
\end{proof}

\medskip
Lemmas \ref{Lemma:SubAdd} and \ref{Lemma:Regularized} 
immediately imply the expression for $\Qca(\cN)$ quoted in Theorem \ref{thm:main}. 

From Theorem \ref{thm:main} we can easily show the following proposition.
\begin{proposition}
  \label{prop:convex}
  $\Qca$ is a convex function of the channel $\cN$.
\end{proposition}
\begin{proof}
Letting $\cN_1$ and $\cN_2$ be channels and $\omega_i = (\id \ox \cN_i \ox \cA_d)\phi$, 
the convexity of $I(A\rangle B\top)_{\omega_{AB\top}}$~\cite{LiebRuskai} gives us
\begin{equation}\nonumber
    I(A\rangle B\top)_{p\omega_1+(1-p)\omega_2}
                \leq pI(A\rangle B\top)_{\omega_1} + (1-p)I(A\rangle B\top)_{\omega_2},
\end{equation}
where $p\omega_1+(1-p)\omega_2 = \bigl[ \id \ox (p\cN_1+(1-p)\cN_2) \ox \cA_d \bigr]\phi$.
This implies  
  \begin{eqnarray}\nonumber
    \max_{\phi} I(A\rangle B\top)_\omega
               &\leq & p\max_{\phi} I(A\rangle B\top)_{\omega_1}\\
& & + (1-p)\max_{\phi} I(A\rangle B\top)_{\omega_2},\nonumber
  \end{eqnarray}
  which tells us exactly that $\Qca\bigl( p\cN_1+(1-p)\cN_2 \bigr)
   \leq p\Qca(\cN_1)+(1-p)\Qca(\cN_2)$.
\end{proof}

\section{Implications for the unassisted capacity}
\label{sec:application}
In this section we explore some of the limitations that
the ss-capacity places on 
the standard capacity of a quantum channel.  
As noted in the introduction, by simply not using the assistance 
channel provided, it is possible to communicate over a 
channel at the unassisted rate.  In other words, 
\begin{equation}
  Q(\cN) \leq \Qca(\cN).
\end{equation}

Furthermore, as we will now see, this upper bound is actually 
an equality for the class of channels known as {\em degradable} \cite{DS03}.
As mentioned above, every channel, $\cN$, can be expressed as an isometry 
$U_{\cN} : A \rightarrow BE$ followed by a partial trace, such that
$\cN(\rho) = \Tr_{E}U_{\cN}\rho U_{\cN}^\dg$. The complementary 
channel of $\cN$, which we call $\widehat{\cN}$, is the channel that 
results by tracing out system $B$ rather than
the environment: $\widehat{\cN}(\rho) = \Tr_{B}U_{\cN}\rho U_{\cN}^\dg$.
Since the Stinespring dilation is unique up to isometric equivalence
of $E$, $\widehat{\cN}$ is well-defined up to isometries on the output.
A channel is degradable if there exists a completely positive 
trace preserving map, $\cD:\cB(B)\rightarrow \cB(E)$, which ``degrades''
the channel $\cN$ to $\widehat{\cN}$.
In other words, $\cD\circ \cN = \widehat{\cN}$.  
The capacity of a degradable channel is given by the single 
letter maximization of the coherent
information, as shown in \cite{DS03}. Furthermore, we will 
now show that the ss-capacity of a degradable channel is given
by the same formula.  That is, the assistance 
channels we have been considering are of no use at all for a degradable channel.

\begin{theorem}\label{Thm:Degradable}
  If $\cN$ is degradable, then $\Qca(\cN) = Q(\cN)$.
\end{theorem}
\begin{proof}
  Fix $\ket{\phi}_{A\tilde{A}W_d}$.  Then, with respect to the state
  $\omega_{AB\top} = (\id_A \ox \cN \ox \cA)\phi$,
  \begin{equation}\label{Eq:DegradeAdd}
    I(A \rangle B\top) \leq I(A\top\bot \rangle B) + I(ABE \rangle \top)
  \end{equation} 
  exactly when $I(E;\perp)\leq I(B;\top)$, which is true if $\cN$ is 
  degradable by the monotonicity of mutual information under local operations (the monotonicity of quantum mutual information
is a special case of the monotonicity of quantum relative entropy, first proved in \cite{Lindblad75}).  
  This implies that the maximum value of the left hand side 
  of Eq.~(\ref{Eq:DegradeAdd}) is no larger than the maximum of the right hand side. 
  The maximum of the first term on the right is exactly
  the single-shot maximization of the coherent information, $Q^{(1)}(\cN)$, 
  whereas the maximum of the second is zero (because of the
  no-cloning argument), so that
  $I(A \rangle B\top)_\omega \leq Q(\cN)$.
  Furthermore, by choosing a trivial assistance channel, the left hand
  side can attain the right hand side.
\end{proof}

\medskip
As an aside, we note that the definition of $\Qcaa$ can be reformulated
in terms of degradable
channels. In particular, we call a channel $\cA :\cB(A)\rightarrow \cB(B)$
with complementary channel $\widehat{\cA} : \cB(A)\rightarrow \cB(E)$
\emph{bidegradable} if both $\cA$ and $\widehat{\cA}$ are degradable, which 
is equivalent to requiring the existence of channels $\cD:\cB(B)\rightarrow \cB(E)$ and
$\cD':\cB(E)\rightarrow \cB(B)$ such that $\cD\circ \cA = \widehat{\cA}$
and $\cD'\circ \widehat{\cA} = \cA$.
Then, using the Stinespring theorem on such $\cA$ and the data processing
inequality for the coherent information (Eq.~(\ref{Eq:DataProcessing})), we have
\[
  \Qcaa(\cN) = \sup_{\cA\text{ bidegradable}} Q^{(1)}(\cN\otimes\cA).
\]

\medskip
Returning to our goal of finding upper bounds for $Q$, we will make use of
Theorem \ref{Thm:Degradable}, which allows us to calculate the 
ss-capacity of any degradable channel.  If a channel $\cN$ 
can be written as a convex combination
of degradable channels, Theorem \ref{Thm:Degradable}, together with the convexity 
of $\Qca$, provides an upper bound for $\Qca(\cN)$ and therefore also $Q(\cN)$.  

For instance, the depolarizing channel can be written as a 
convex combination of dephasing-type channels, 
\begin{eqnarray}\nonumber
  \cN_p(\rho) &=& (1-p)\rho + \frac{p}{3}X\rho X
                            + \frac{p}{3}Y\rho Y+ \frac{p}{3}Z\rho Z\\
            \nonumber  &=& \frac{1}{3}\cX_p(\rho) +  \frac{1}{3}\cY_p(\rho)
                            + \frac{1}{3}\cZ_p(\rho), 
\end{eqnarray}
where $\cX_p(\rho) = (1-p)\rho + pX\rho X$ and 
similarly for $\cY_p$ and $\cZ_p$.  From this we conclude that 
\begin{equation} \nonumber
\Qca(\cN_p) \leq \frac{1}{3}\Qca(\cX_p)
                   +\frac{1}{3}\Qca(\cY_p)+\frac{1}{3}\Qca(\cZ_p)
            =    1 - H(p),
\end{equation}
where we have used the fact that $\cX_p$, $\cY_p$, and $\cZ_p$ are degradable 
and have ss-capacity $1-H(p)$ (Theorem \ref{Thm:Degradable}).
This reproduces the upper bounds of \cite{VP97,Rains99,Rains-PPT},
which have been the best known for small $p$.
  
We can also evaluate $\Qca(\cN_p)$ for 
$p=\frac{1}{4}$ as follows.  For this value of $p$, there 
is a CP-map which can be composed with 
the complementary channel, $\widehat{\cN}_p$, to generate $\cN_p$ \cite{BDEFMS98}.  
This immediately implies $\Qca(\cN_{1/4})=0$, since otherwise both 
Bob and Eve could both reconstruct
the encoded state with high fidelity, giving a 
violation of the no-cloning theorem.  More explicitly, 
for any state $\ket{\phi}_{A\tilde{A}\top\perp}$ with the
symmetry $\top\leftrightarrow\perp$ we have, with respect to
the state $(\id\ox\cN_{1/4})\phi$,
\begin{equation}
  I(A\rangle B\top) = -I(A\rangle E\top) \leq -I(A\rangle B\top),
\end{equation}
from which  we conclude $\Qca(\cN_{1/4})=0$, and where 
the second step is due to the quantum data processing inequality  (Eq.~(\ref{Eq:DataProcessing})).
This reproduces the bound of \cite{BDEFMS98}, and 
furthermore, because the ss-capacity is convex,
we find that
\begin{equation}
  Q(\cN_p) \leq \Qca(\cN_p)
                     \leq {\rm conv}\bigl( 1-H(p), (1-4p)_+ \bigr),
\end{equation}
with the notation
\begin{equation*}
  x_+ = \begin{cases}
          x & \text{ if } x\geq 0, \\
          0 & \text{ if } x<0.
        \end{cases}
\end{equation*}

It is important to note that the quantum capacity $Q$ 
is not known to be convex and, indeed, may well not be---in the two way
scenario, both nonadditivity and nonconvexity would be implied \cite{SST00} by the conjecture
of  \cite{DSSTT00} that a family of Nonpositive Partial Transpose (NPT) Werner states is bound entangled.  
Thus, while the two bounds above were already known, it was 
not clear that the convex hull of these was also an upper bound.

\begin{figure}[htbp]
  \centering
  \hspace{1.5cm}
  \includegraphics[width=8cm]{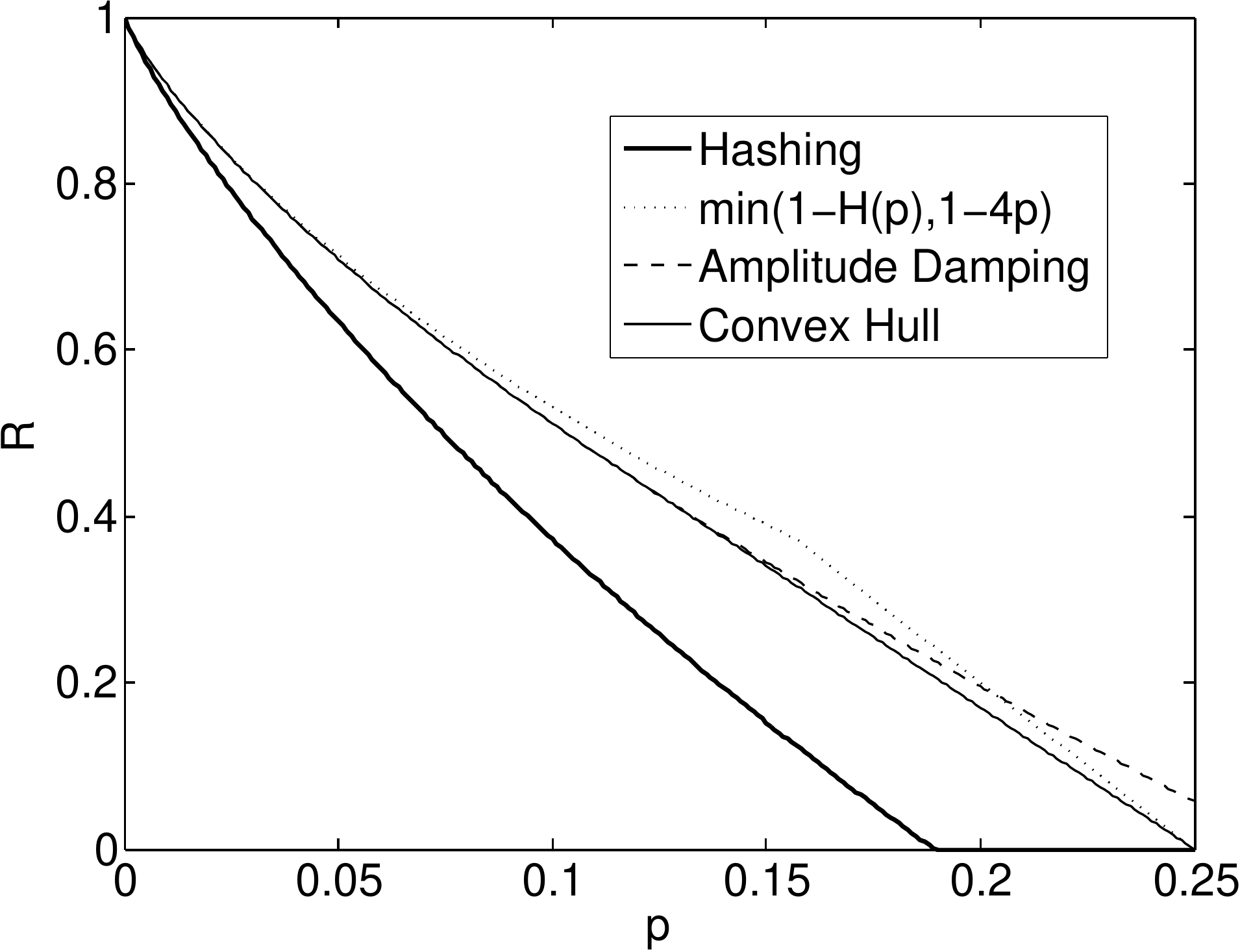}
  \hspace{1.5cm}
  \caption{Our upper bound evaluated for the depolarizing channel:
  the dotted line is the previous best bound that comes from the minimum of a no-cloning argument and Rains' bound,
  the dashed line is the capacity of an amplitude damping
  channel with damping parameter $\gamma_p = 4\sqrt{1-p}(1 - \sqrt{1-p})$;
  finally, the thin solid line is the convex hull of the first two,
  our best upper bound on $\Qca(\cN_p)$ and $Q(\cN_p)$ so far;
  The thick solid line is the hashing (lower) bound, $1-H(p)-p\log 3$.}
  \label{figure:Graph}
\end{figure}

We will now provide a tighter bound for $\Qca(\cN_p)$, by 
expressing the depolarizing channel as a convex combination
of amplitude-damping channels, which were shown to be degradable
in \cite{GF04}.  The amplitude-damping channel can be expressed as
\begin{equation}
  \Delta_{\gamma}(\rho) = A_0\rho A_0^\dagger + A_1\rho A_1^\dagger,
\end{equation}
where
\begin{equation}
  A_0 = \left(\begin{matrix} 1 & 0 \\
                             0 & \sqrt{1-\gamma}
              \end{matrix}\right)\quad \text{ and }\quad \\
  A_1 = \left(\begin{matrix} 0 & \sqrt{\gamma} \\ 
                             0 & 0\end{matrix}\right).
\end{equation}
From this we find that
\begin{equation} \nonumber
\frac{1}{2}\Delta_{\gamma}\left(\rho\right)
        + \frac{1}{2}Y\, \Delta_{\gamma}\left(Y\rho Y\right)\, Y = \cN_{(q,q,p_z)}(\rho),
\end{equation}
where
\begin{equation}\nonumber
  \cN_{(q,q,p_z)}(\rho) = \left(1-2q-p_z\right)\rho + q X\rho X + q Y \rho Y + p_zZ\rho Z,
\end{equation}
with $q = \frac{\gamma}{4}$ and
$p_z = \frac{1}{2}\left(1 - \frac{\gamma}{2}- \sqrt{1-\gamma}\right)$.
The depolarizing channel can now be expressed as
\begin{equation}
\cN_{2q+p_z} = \frac{1}{3}\cN_{(q,q,p_z)}
                       +\frac{1}{3}\cN_{(q,p_z,q)} + \frac{1}{3}\cN_{(p_z,q,q)},
\end{equation}
so that $\cN_{p}$ is a convex combination 
of amplitude damping channels with $\gamma_p = 4\sqrt{1-p}\left(1-\sqrt{1-p}\right)$.
This gives us an upper bound, shown in Figure \ref{figure:Graph}, of 
\begin{equation}
  Q(\cN_p) \leq \Qca(\cN_p)
           \leq {\rm conv}\bigl( Q(\Delta_{\gamma_p}),(1-4p)_+ \bigr),
\end{equation}
where $Q(\Delta_{\gamma_p})$ is, according to \cite{GF04}, given by
\begin{equation}
  Q(\Delta_{\gamma_p}) = \max_{0\leq t\leq 1}
                           \Bigl[ H_2\bigl(t(1-\gamma_p)\bigr)-H_2(t\gamma_p) \Bigr].
\end{equation}
The resulting bound is strictly stronger than the previously known bounds of $1-H(p)$ and $(1-4p)_+$ for all $0.25~>~p~>~0.04$.

\section{A lower bound for $Q_{ss}$}
\label{sec:CA-LB}
\begin{figure}[htbp]
  \centering

  \includegraphics[width=8cm]{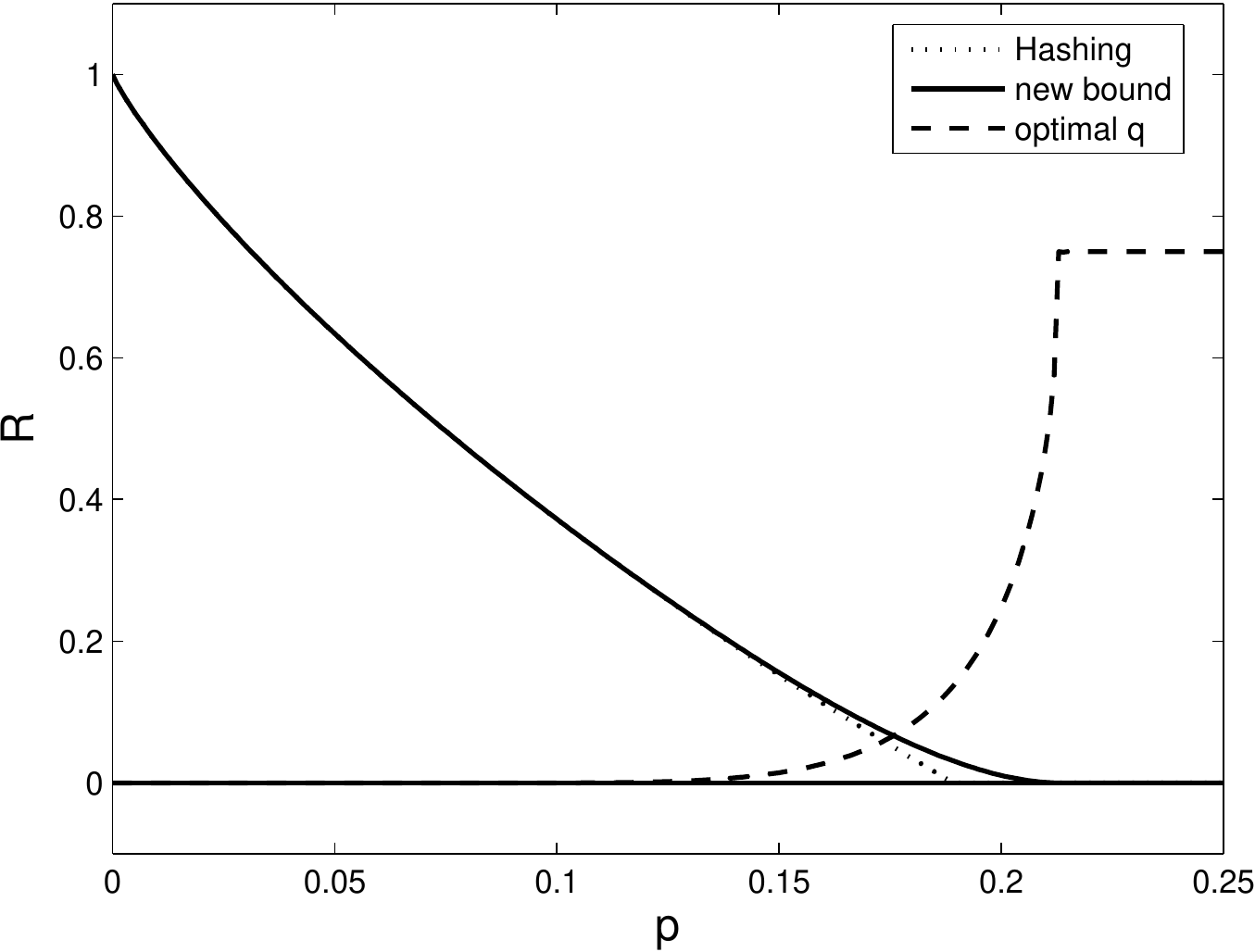}
  
  \caption{Our lower bound for the symmetric side channel capacity of the depolarizing channel: 
The dotted curve is the hashing lower bound for $Q_{ss}$, which in this case is $1-H(p)-p\log 3$.  
The solid curve is Eq~(\ref{eq:DepLB}), evaluated for the optimal value of $q$.  The dashed curve is the optimal value of $q$.}

  \label{figure:LowerBound}
\end{figure}

In this section we present a particular state relative to which the quantity optimized in 
Eq.~(\ref{eq:asym}) to give $Q_{ss}$ is, for the depolarizing channel, strictly larger than the hashing lower bound for $Q_{ss}$
mentioned in the previous section.  Letting 
\begin{equation}\label{eq:DepolarState}
\ket{\phi} = \sum_{s,t =0}^1\sqrt{q_{st}}X^sZ^t\ox \1 \ket{\Phi^+}_{A\tilde{A}}\ket{s t}_{F},
\end{equation}
we have
\begin{eqnarray}
 \Qcaa(\cN) &=& \sup_{\rho_{A\tilde{A}F}}
                    \frac{1}{2}\bigl[ I(A\rangle BF) - I(A\rangle EF) \bigr]\\
 & \geq &  \frac{1}{2}I(A\rangle BF)_{(\id_{AF}\ox \cN_p)(\phi)} \nonumber\\
& & +\frac{1}{2}I(A\rangle B)_{(\id_{AF}\ox \cN_p)(\phi)}\nonumber
\end{eqnarray}
for any choice of $q_{st}$ with $\sum_{st}q_{st} =1$.  For the depolarizing channel, the optimal such $q_{st}$ is of the form
\begin{equation}\label{eq:Qsymmetry}
q_{st} = (1-q,q/3,q/3,q/3),
\end{equation} 
which leads to entropies
\begin{eqnarray}\nonumber
S(BF) &=& -\left[\frac{1}{2}-\frac{4pq}{9} - 2\eta_{p,q}\right]\log\left[ \frac{1}{4}-\frac{2pq}{9} - \eta_{p,q}\right]\\
& & -\left[\frac{1}{2}-\frac{4pq}{9}+2\eta_{p,q}\right]\log\left[\frac{1}{4}-\frac{2pq}{9}+\eta_{p,q}\right]\nonumber \\
& & - \frac{8pq}{9}\log\left[\frac{2pq}{9}\right] \label{Eq:SBF}
\end{eqnarray}

\begin{eqnarray}
S(AB) & = & -\left[1-p-q +\frac{4pq}{3}\right]\log\left[1-p-q +\frac{4pq}{3}\right] \nonumber\\
& & -\left[p+q-\frac{4pq}{3}\right]\log \left[\frac{p+q}{3}-\frac{4pq}{9}\right]\label{Eq:SAB}
\end{eqnarray}

\begin{eqnarray}
S(B)& = & 1\nonumber\\
S(ABF)& = & H(p)+p\log 3,\nonumber
\end{eqnarray}
where 
\begin{equation}
\eta_{p,q} = \frac{1}{36}\sqrt{81-720pq-512p^2q^2+576qp(p+q)}.
\end{equation}
This gives a lower bound of 
\begin{eqnarray}
\Qca(\cN) & \geq & \frac{1}{2} \left( 1 - H(p)- p\log 3\right) \nonumber\\
& & + \frac{1}{2}\left( S(BF)-S(AB)\right) \label{eq:DepLB}, 
\end{eqnarray}
with $S(BF)$ and $S(AB)$ given by Eqs.~(\ref{Eq:SBF}) and (\ref{Eq:SAB}), respectively.  This, 
optimized over $q$, is plotted in  Fig.~\ref{figure:LowerBound}.
The resulting bound is nonzero up to $p = 0.2124$, which should be compared to the threshold of hashing at $p=0.1893$ and of the best known codes for 
the depolarizing channel at $0.19088$ \cite{SmithSmo0506}.

It is intriguing that the form of Eq.~(\ref{eq:DepolarState}) corresponds to a preprocessing 
of $\cN_p$'s input by applying a depolarizing channel whose environment is $F$, then sending 
$F$ through the side channel, with the optimal level of preprocesssing noise increasing to the completely depolarizing probability of
 $3/4$ as $\cN_p$'s noise level increases.

\section{One-way distillation with symmetric side channels}
\label{sec:ca-1-way}
Based on the connection between quantum channel capacities and entanglement distillation via
local operations with one-way classical communication (1-LOCC) \cite{BDSW96,BarnumKN00},
we can define a symmetric side channel assisted distillation notion for
bipartite states $\rho_{AB}$:
\begin{equation}
  \label{eq:Dcaa}
  \Dcaa(\rho) = \sup_{\sigma, \cE}
                      I(A^\prime \rangle B\widetilde{B})_{(\cE\ox \id_{B\tilde{B}})\rho\otimes\sigma},
\end{equation}
where the supremum is over states $\sigma_{\tilde{A}\tilde{B}\tilde{E}}$ (such that $\tilde{B} \simeq \tilde{E}$)
with the property $\sigma_{\tilde{A}\tilde{B}} = \sigma_{\tilde{A}\tilde{E}}$ and operations on Alice's 
system $\cE:A\tilde{A}\rightarrow A^\prime$.
Observe that these states (or rather their restrictions $\sigma_{\tilde{A}\tilde{B}}$)
are often called \emph{two-shareable} or {\em two-extendable} in the
literature. Note also that without loss of generality we may restrict our attention to pure states,
at the expense of increasing the dimension of their local supports
(which, in any case, is unbounded in the above definition).

For a state $\rho_{AB}$ with purification
$\ket{\phi}_{ABE}$ and with respect to the state $\omega_{A'BEF} = (\cT_A\ox\id_{BE})\phi$, with $\cT:\cB(A)\rightarrow \cB(A'\ox F)$
we have the analogue of Lemma~\ref{lemma:asymmetric}:
\begin{equation}
  \label{eq:Dca-original}
  \Dcaa(\rho) = \sup_{\cT:A\rightarrow A'F}
                 \frac{1}{2}\bigl( I(A'\rangle BF) - I(A'\rangle EF) \bigr).
\end{equation}

Just as for channels, we find that $\Dcaa$ is additive, convex
and indeed a 1-LOCC entanglement monotone, reducing to the
entropy of entanglement for pure states, and vanishing for all
two-shareable states. Furthermore, $\Dcaa(\rho)$ has an operational meaning---it is the
one-way distillable entanglement of $\rho$ when assisted by arbitrary two-shareable states.

The notion of degradability of channels is translated to states as
follows: $\rho_{AB}$ is called \emph{degradable} if, for its
purification $\phi_{ABE}$, there exists a quantum channel
$\cD:\cB(B) \rightarrow \cB(E)$ such that $\phi_{AE} = (\id_A\ox\cD)\rho_{AB}$.
The analogue of the bidegradable channels are states $\sigma_{ABE}$
such that there are channels degrading both ways, $B\rightarrow E$
and $E\rightarrow B$.

Analogously to our findings for channels, we can prove that
$\Dca(\rho) = D_\rightarrow(\rho)$ for degradable states, so that
the upper bounds in the previous section on the quantum capacity
of the depolarizing channels, including Fig.~\ref{figure:Graph},
translate into upper bounds on the one-way distillable entanglement
of two-qubit Werner states.

\section{Quantum value added}
\label{sec:value}

In Section \ref{sec:application} we saw that the ss-capacity of a degradable channel is
equal to its unassisted capacity.  In fact, we have not been able to show a separation between the ss-capacity and 
the unassisted capacity for {\em any} channel.  The question arises: Are there $\cN$ such that $\Qca(\cN)>Q(\cN)$?

Motivated by this question,  for any CPTP map $\cM$, we define the {\em value added} of $\cM$ to be
\begin{equation}\label{eq:DefValueAdded}
V^{(1)}(\cM):= \sup_{\cN}\left[Q^{(1)}(\cN\ox\cM)- Q^{(1)}(\cN)\right].
\end{equation}

In words, $V^{(1)}(\cM)$ is the largest increase in the optimized coherent information that $\cM$ can provide when 
used as a side channel for some other $\cN$.  This definition has the appealing property that $V^{(1)}$ is 
sub-additive, since

\begin{equation}
V^{(1)}(\cM_1\ox \cM_2) \ \ \ \ \ \ \ \ \ \ \ \ \ \ \ \ \ \ \ \ \ \ \ \ \ \ \ \ \ \ \ \ \ \ \ \ \ \ \ \ \ \ \ \ \ \ \ \  \nonumber\\
\end{equation}
\vspace{-0.5cm}
\begin{eqnarray}
&=& \sup_{\cN}\left[Q^{(1)}(\cN\ox\cM_1\ox\cM_2)- Q^{(1)}(\cN)\right]\nonumber\\
&\leq&  \sup_{\cN}\left[Q^{(1)}(\cN\ox\cM_1\ox\cM_2)- Q^{(1)}(\cN\ox \cM_2)\right] \nonumber\\
& & + \sup_{\cN}\left[Q^{(1)}(\cN\ox \cM_2)- Q^{(1)}(\cN)\right]\nonumber\\
 &\leq&  V^{(1)}(\cM_1)+ V^{(1)}(\cM_2).\nonumber
\end{eqnarray}
Letting 
\begin{equation}
V(\cM) := \lim_{n \rightarrow \infty}\frac{1}{n}V^{(1)}(\cM^{\ox n}),\nonumber
\end{equation}
we have $V(\cM)\leq V^{(1)}(\cM)$, and furthermore, for all $\e>0$ and sufficiently large $n$ 
\begin{eqnarray}
 V^{(1)}(\cM^{\ox n}) & = &  \sup_{\cN}\left[Q^{(1)}(\cN\ox\cM^{\ox n})- Q^{(1)}(\cN)\right]\nonumber\\
& \geq & \left[Q^{(1)}(\cM^{\ox n}\ox\cM^{\ox n})- Q^{(1)}(\cM^{\ox n})\right]\nonumber\\
& \geq & (2n)\left(Q(\cM)-\e\right) - nQ(\cM),\nonumber
\end{eqnarray}
so that
\begin{eqnarray}
\frac{1}{n}V^{(1)}(\cM^{\ox n}) & \geq & Q(\cM) - 2\e,\nonumber
\end{eqnarray}
which gives us $V^{(1)}(\cM) \geq V(\cM) \geq Q(\cM)$.

In addition to this upper bound for the capacity, $V^{(1)}$ also provides a sufficient condition for $\Qca(\cN)=Q(\cN)$:
\begin{equation}\nonumber
Q_{ss}(\cN)-Q(\cN) \ \ \ \ \ \ \ \ \ \ \ \ \ \ \ \ \ \ \ \ \ \ \ \ \ \ \ \ \ \ \ \ \ \ \ \ \ \ \ \ \ \ \ \ \ \ \ \ \ \ \ \ \ 
\end{equation}
\vspace{-0.6cm}
\begin{eqnarray}
& = & \lim_{n\rightarrow \infty}\frac{1}{n}\left(\sup_{d}Q^{(1)}(\cN^{\ox n}\ox \cA_d) - Q^{(1)}(\cN^{\ox n})\right)\nonumber\\
& \leq & \lim_{n\rightarrow \infty}\frac{1}{n}\left(\sup_{d}\sup_{\cM}\left(Q^{(1)}(\cM \ox \cA_d) - Q^{(1)}(\cM)\right)\right)\nonumber\\
& \leq & \sup_{d}V^{(1)}(\cA_d),\nonumber
\end{eqnarray}
so that $Q_{ss}(\cN) = Q(\cN)$ for all $\cN$ as long as $V^{(1)}(\cA_d) = 0$ for all $d$.  Unfortunately, although Eq.~(\ref{eq:DefValueAdded}) is
nominally single-letter, evaluating $V^{(1)}$ seems to be quite difficult, as it contains an optimization over an infinite number of variables.

\section{Discussion}
\label{sec:discussion}
We have studied the capacity of a quantum channel given the assistance 
of an arbitrary symmetric side channel.
The capacity formula we find is in many ways more manageable than the 
known expression for the (unassisted) 
quantum capacity, and we are able to establish that the ss-capacity 
is both convex and additive.  
By taking advantage of the convexity of $\Qca$ and the fact 
that $\Qca$ and $Q$ coincide for degradable channels, 
we presented a general method for finding upper bounds to $Q$ 
and in particular provided a bound for the capacity 
of the depolarizing channel that is stronger than any previously 
known result.

We have left many questions unanswered.
The most pressing is whether it is possible to bound 
the dimension of the symmetric side channel needed to achieve 
the ss-capacity.  Such a bound would allow us to evaluate 
$\Qca(\cN)$ efficiently, which we expect would provide very tight
bounds on $Q$ in many cases.

So far, we have not been able to find a channel for which 
the ss-capacity and capacity differ.  We expect that such channels exist, and 
a better understanding of when the two capacities differ may 
point towards simplifications of the quantum capacity 
formula in Eq.~(\ref{Eq:QuantumCapacity}).

It is worth mentioning that we first discovered the 
unsymmetrized version of the quantity $\Qcaa$ given in Lemma~\ref{lemma:asymmetric},
and that it is an upper bound for $Q$. This was motivated by
the quest to find the entanglement analogue of the upper 
bound on distillable key presented in \cite{KGR05,RGK05}.  
It was only later that it became clear that the formula could be made symmetric 
and interpreted as the quantum capacity of a channel 
given the family of assistance 
channels we have considered.

Finally, it should be noted that the approach we have taken here 
is qualitatively similar to the work of \cite{VP97,Rains99,Rains-PPT}
in the two-way scenario.  In that work, it was found that {\em enlarging} the
set of operations allowed for entanglement distillation from LOCC 
to the easier-to-deal-with set of separable or positive-partial-transpose-(PPT-)preserving 
operations made it possible to establish tighter bounds on two-way distillable entanglement
than was possible by considering LOCC protocols directly.  
Similarly, we have shown that by augmenting a channel
with a zero capacity side channel,
a simplified capacity formula can be found 
that allows us to establish tighter bounds on the unassisted 
capacity than were possible by 
direct considerations.  To what extent this approach can be
used in general, the reason such an approach works at all, and the tightness of 
the bounds achieved in this way are all questions that we leave wide open.

\section*{acknowledgments}
It is a pleasure to thank Andrew Childs, Mary-Beth Ruskai, and Frank Verstraete for
illuminating conversations about degradable channels and symmetric assistance.

\section*{Biographies}

Graeme Smith received the B.Sc. degree in physics from the University of Toronto, Toronto, ON, Canada, 
in 2001 and the M.S. and Ph.D. degrees in physics from the California Institute of Technology, Pasadena
in 2004 and 2006, respectively.  

He is currently a Postdoctoral Fellow at the IBM T.J. Watson 
Research Center, Yorktown Heights, NY, working on quantum information theory, coding theory, 
and cryptography.

John A. Smolin received the S.B. degree in physics from the Massachusetts
Institute of Technology (MIT), Cambridge, in 1989 and the Ph.D. degree, 
also in physics from the University of California, Los Angeles, in 1996.

After receiving the Ph.D. degree, he has been at IBM T.J. Watson 
Research Center, Yorktown Heights, NY, first and a postdoc and 
subsequently as a Research Staff Member.  He, together with Charles 
Bennett built the first quantum cryptography apparatus at IBM in 1989.
His current research interests are in quantum information theory, coding theory,  
and quantum computation, with the occasional misguided foray into the foundations of quantum mechanics.

Andreas Winter was born in Muhldorf am Inn, Germany in 1971.  He received 
the Diploma degree in mathematics from the Freie Universitat Berlin, 
Berlin, Germany, in 1997.  In 1999 he received the Ph.D. degree from the
Fakultat fur Mathematik, Universitat Bielefeld, Bielefeld, Germany. 

He was a Research Assistant at the University of Bielefeld until
2001, and since there has been with the University of Bristol, Bristol, U.K., 
most recently as Professor of Mathematics.  His research interests
include quantum information theory, complexity theory, and discrete mathematics.
He is currently Associate Editor for Quantum Information Theory for the IEEE Transactions
on Information Theory.

\bibliographystyle{IEEEtran}

\end{document}